\documentclass[journal]{IEEEtran}

\usepackage{cite}
\usepackage{array}
\usepackage{arydshln}
\usepackage{comment}
\usepackage{amssymb}
\usepackage{kotex}

\ifCLASSINFOpdf
  \usepackage[pdftex]{graphicx}
\else
\fi

\usepackage {xcolor}
\usepackage{amsmath,amsthm,amsfonts,amscd}  

\usepackage{algorithmic}
\usepackage{algorithm}

\usepackage{array}
\usepackage{multirow}

\ifCLASSOPTIONcompsoc
  \usepackage[caption=false,font=normalsize,labelfont=sf,textfont=sf]{subfig}
\else
  \usepackage[caption=false,font=footnotesize]{subfig}
\fi

\usepackage{url}

\begin{document}
%
\title{Numerical Evaluation of Angle-Dependent IR-Transparent Radiative Cooling Performance for Asymmetric Periodic Structures}
%
%
%

\author{Junwoo~Gim, 
Jun~Heo, 
Weng Cho Chew, and
Dong-Yeop Na,
\thanks{Junwoo~Gim, Jun Heo, and Dong-Yeop Na are with the Department of Electrical Engineering, Pohang University of Science and Technology, Pohang, Gyeongsangbuk-do 37673, South Korea (e-mail: dyna22@postech.ac.kr). Weng C. Chew is with the Elmore Family School of Electrical and Computer Engineering, Purdue University, West Lafayette, IN 47907, USA.}}

%
%

\markboth{Preprint}%
{Shell \MakeLowercase{\textit{et al.}}: Bare Demo of IEEEtran.cls for IEEE Journals}
%

\maketitle

\begin{abstract}
Infrared (IR)-transparent passive radiative cooling (PRC) enables non-contact thermal management by regulating radiative heat exchange without direct attachment to the cooling object.
While asymmetric IR transmission at a specific incidence angle---typically normal incidence---is often emphasized, we show that such single-angle asymmetry is neither sufficient nor predictive of practical cooling performance.
In this work, we demonstrate that effective non-contact PRC requires angularly distributed asymmetric IR transparency evaluated through hemispherical integration over emission directions, rather than asymmetry at a single incidence angle.
To quantify this effect, an angle-resolved full-wave electromagnetic (EM) model with Bloch periodic boundary conditions and Floquet mode decomposition is employed to compute wavelength- and angle-dependent bidirectional reflection and transmission of periodic PRC structures.
The resulting EM response is coupled to an energy-balance-based thermal model to predict the transient temperature evolution of the cooling object.
By comparing models that account for the full angular distribution with normal-incidence-only approximations, we show that pronounced asymmetric transmission at normal incidence is generally not preserved at oblique angles.
As a result, angular integration yields only marginal cooling or may even result in net heating, whereas normal-incidence-based models can substantially overestimate cooling performance.
These results establish angularly distributed asymmetric transparency as a key EM design principle for IR-transparent PRC and wide-angle asymmetric metasurfaces.
\end{abstract}


\IEEEpeerreviewmaketitle

\section{Introduction}
\label{sec:Introduction}
\IEEEPARstart{P}{assive} radiative cooling (PRC) enables sub-ambient temperature reduction without external power consumption by reflecting solar radiation while emitting thermal energy from a cooling object to the environment~\cite{raman2014passive}.
Conventional PRC implementations typically rely on solar-reflective coatings combined with mid-infrared (mid-IR) emissive layers operating within the $8$--$13~\mu\mathrm{m}$ atmospheric window.
More recently, infrared-transparent (IR-transparent) PRC concepts have been introduced to address scenarios in which direct thermal contact with the cooling object is impractical, such as plant radiative cooling and textile-based personal cooling~\cite{tong2015infrared}.
In these approaches, an IR-transparent window allows object-emitted thermal radiation to escape while blocking incident solar radiation.
However, because many IR-transparent materials are inherently bidirectional to thermal radiation, inward penetration of ambient IR can significantly degrade cooling performance under realistic environmental conditions.
\begin{figure}[t]
    \centering    
    \includegraphics[width=.8\linewidth]{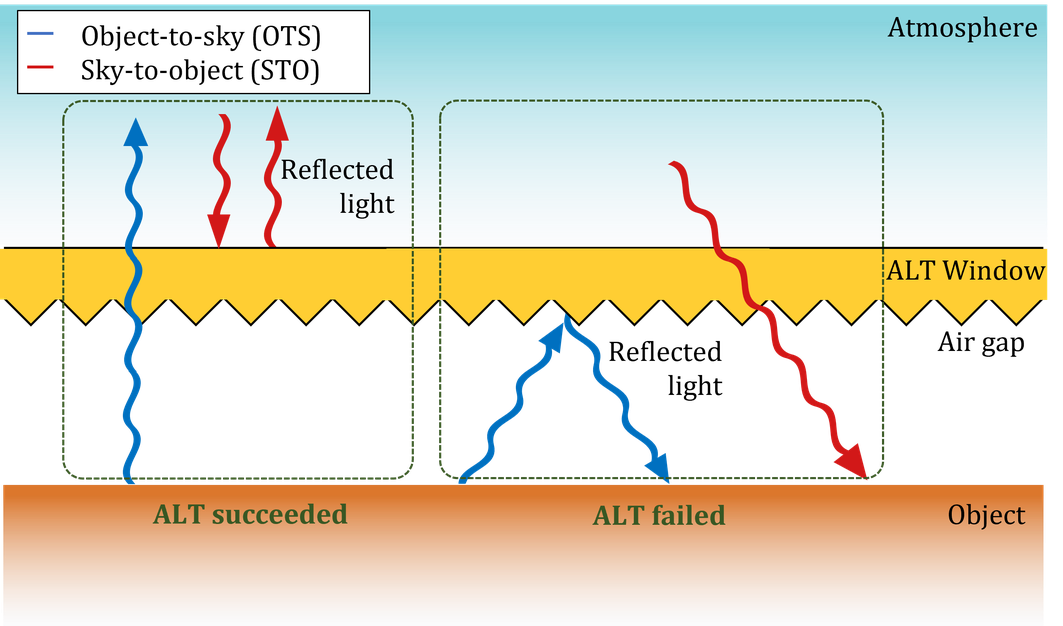}    
    \caption{Schematic of the AEMT device equipped with the ALT function.}    
    \label{fig:AEMT}
\end{figure}

To mitigate this limitation, ultra-broadband asymmetric-emissive/transparent (AEMT) structures have been proposed to enable the asymmetric light transmission (ALT) physics, more specifically, direction-dependent IR transport, suppressing ambient thermal radiation in the sky-to-object (STO) direction while preferentially transmitting thermal emission from the cooling object~\cite{wong2018ultra}.
Various metasurface-, metamaterial-, and nanophotonic-based designs have reported pronounced asymmetric transmission under normal incidence~\cite{shen2015broadband,pfeiffer2014high,ozer2018broadband,ly2022dual,sheikh2023asymmetric}.
However, asymmetric transmission observed at a single incidence angle does not, in general, guarantee effective radiative cooling.
As illustrated in Fig.~\ref{fig:AEMT}, such asymmetry is often not preserved at oblique angles, leading to increased transmission in the STO direction and reduced transmission in the object-to-sky (OTS) direction.

More broadly, AEMT structures exemplify a class of non-contact PRC systems in which the intermediate layer does not act as a heat sink through absorption or emission, but instead modulates radiative heat exchange by redistributing angular and spectral transport channels.
An IR-transparent and effectively lossless PRC layer can therefore engineer asymmetric radiative transfer without violating the second law of thermodynamics.
Because thermal radiation is inherently distributed over a half-space, net cooling performance is governed by angularly weighted integration over emission directions rather than by transmission characteristics at a single incidence angle.
As a result, asymmetric transmission confined to a limited angular range may yield negligible cooling, whereas angularly distributed asymmetry can produce a net imbalance between outgoing and incoming thermal radiation.

In this work, we investigate the role of angularly distributed asymmetric IR transparency in determining the cooling performance of AEMT-based PRC structures.
A proof-of-concept periodic triangular unit cell, adopted as a representative benchmark structure~\cite{Retsch2022}, is evaluated to examine how asymmetric reflection and transmission characteristics---well pronounced at normal incidence---evolve toward grazing incidence and influence practical PRC performance.
The unit cell is analyzed using an angle-resolved full-wave electromagnetic (EM) model with Bloch periodic boundary conditions (PBC), implemented within the discrete exterior calculus (DEC) framework~\cite{teixeira1999lattice,he2007differential,kim2011parallel,teixeira2013differential,chen2017electromagnetic,na2019finite,zhang2023Aphi}, where the bidirectional reflection ($R$) and transmission ($T$) are extracted through Floquet mode decomposition.
The resulting EM response is incorporated into an energy-balance-based thermal model to predict the transient temperature evolution of the cooling object.
By comparing full angular models with normal-incidence-only approximations, we demonstrate that single-angle asymmetry can substantially overestimate cooling performance.
Our results establish angularly distributed asymmetric transparency as a key EM design principle for wide-angle metasurface-based radiative cooling.

\begin{figure}[t]
    \centering        
    \subfloat[Test structure (3D view)\label{fig:Test_Structure}]{
        \includegraphics[height=4.2cm]{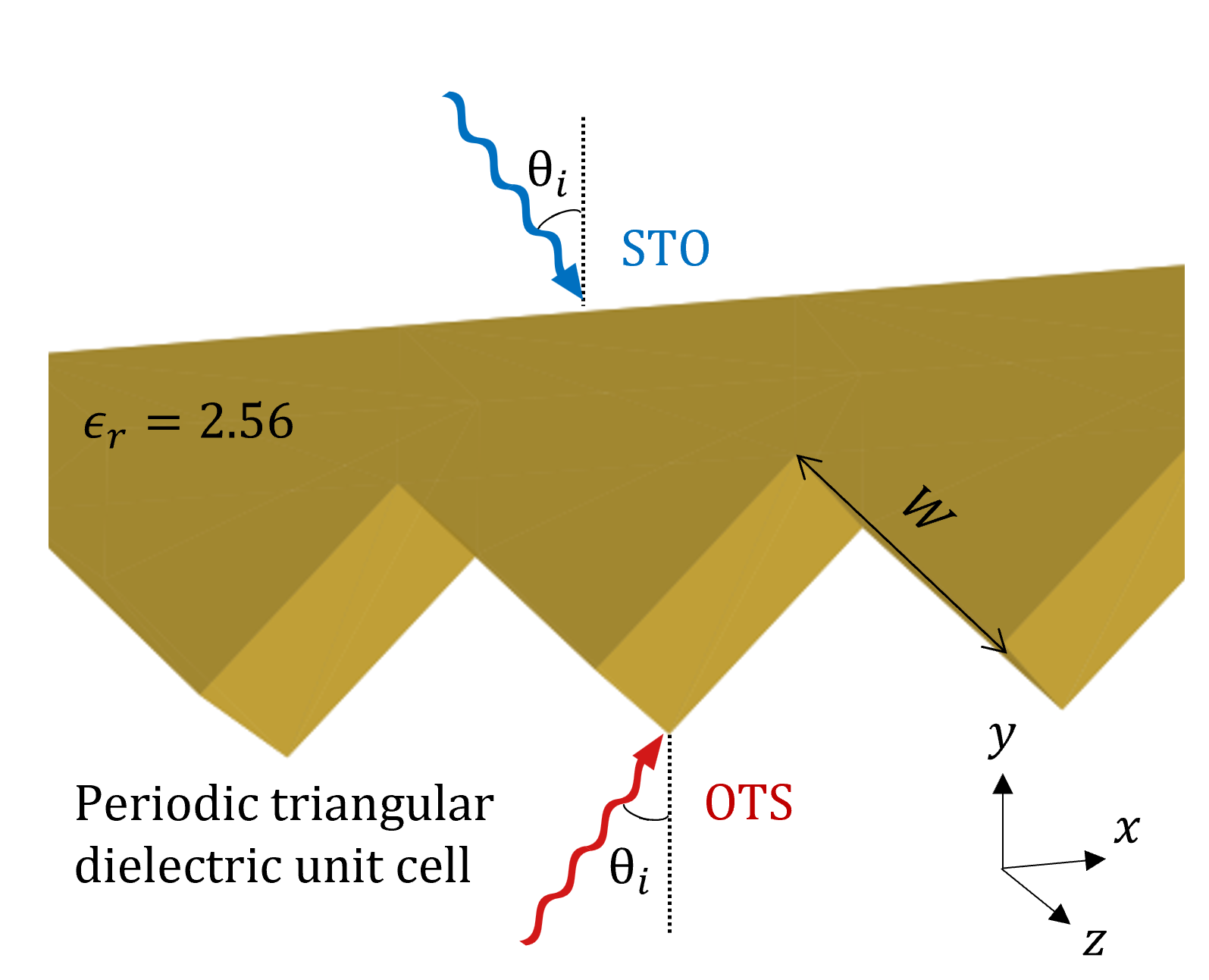}
    }    
    \subfloat[Unit cell\label{fig:2D_Unitcell}]{
        \includegraphics[height=4.2cm]{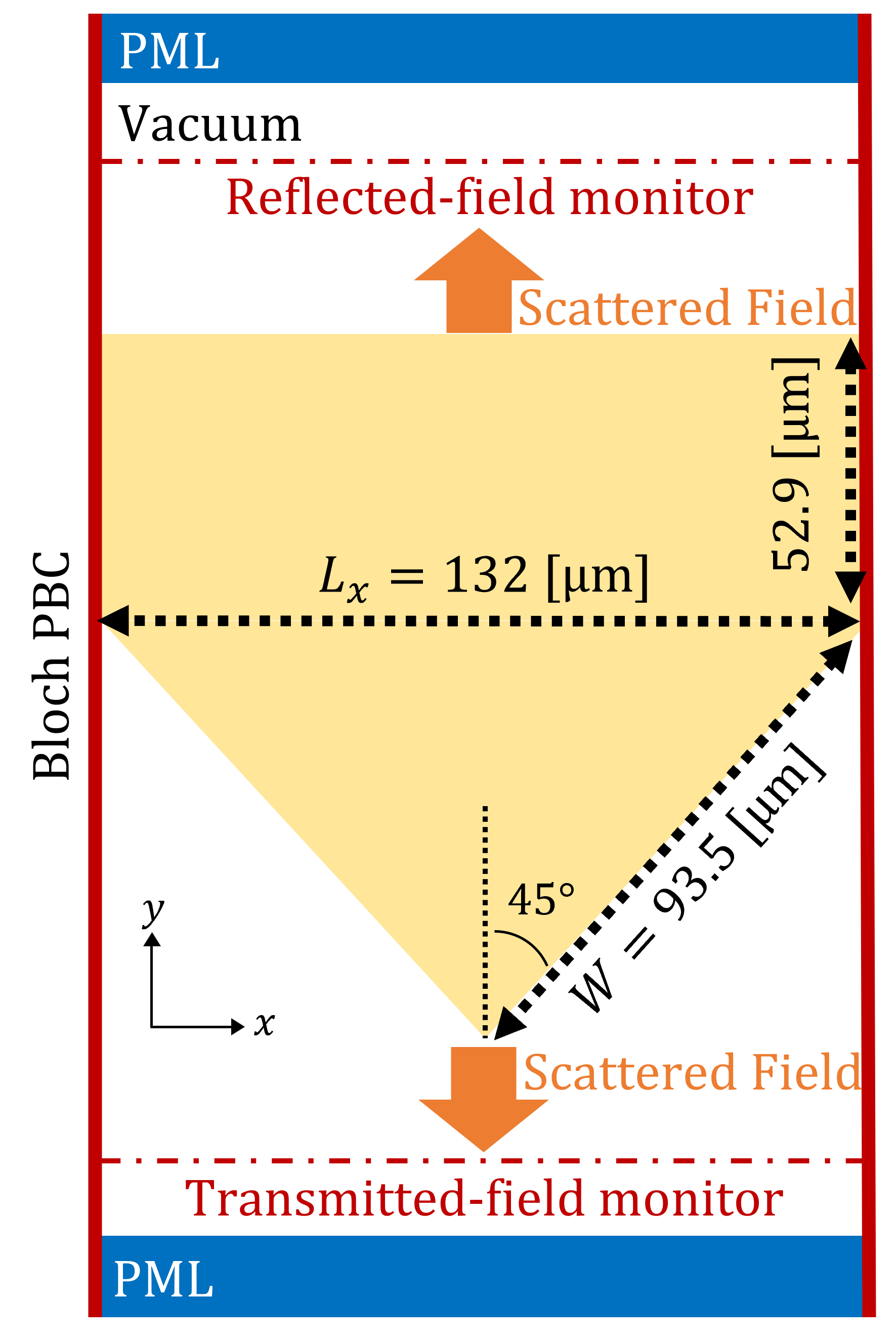}
    }    
    \caption{
    (a) Corresponding test structure, and (b) the 2D periodic unit cell employed for numerical analysis with Bloch PBC.
    }    
    \label{fig:AEMT_Structure_Unitcell}
\end{figure}

\begin{figure}[t]
\centering
\includegraphics[width=.9\linewidth]{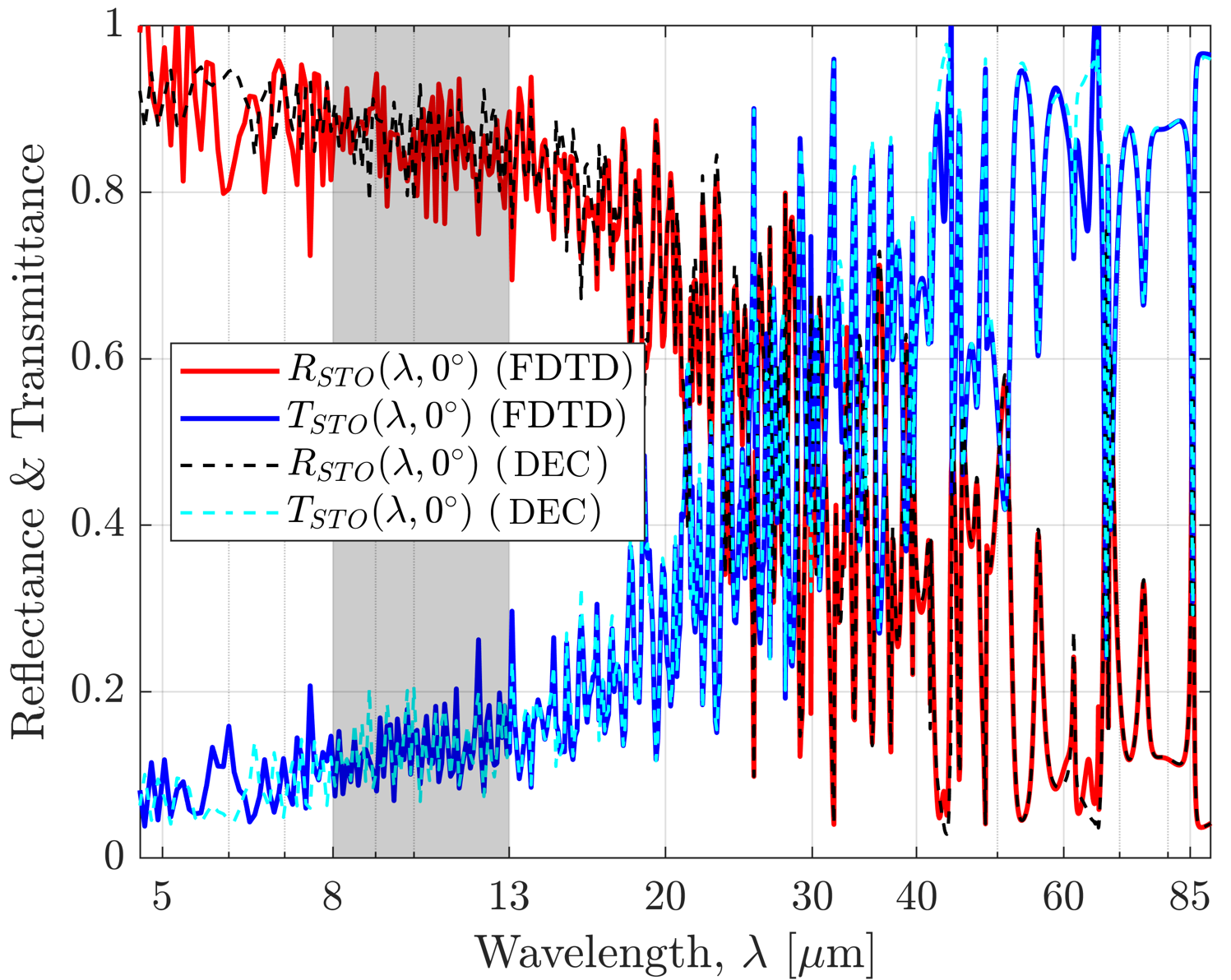}
\caption{Comparison of reflectance (R) and transmittance (T) spectra of the AEMT structure obtained using the FDTD and the in-house DEC solvers under normal-incidence conditions.
The agreement between the two results verifies the accuracy of the developed DEC solver.}
\label{fig:femfdtd}
\end{figure}

\begin{figure*}[t]
    \centering
    \subfloat[Reflectance (STO)\label{fig:2a}]{
        \includegraphics[width=0.24\textwidth]{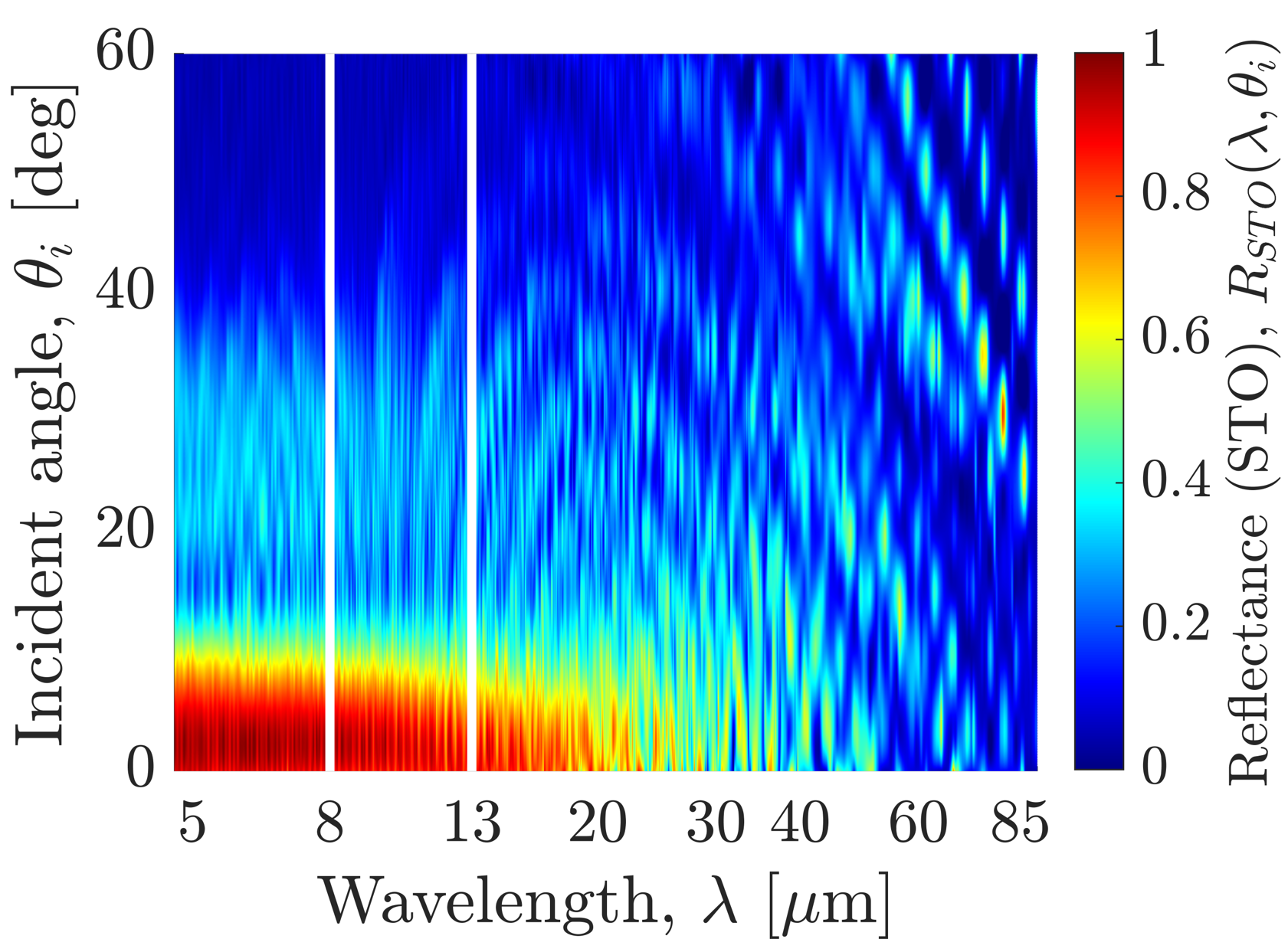}
    }
    \subfloat[Transmittance (STO)\label{fig:2b}]{
        \includegraphics[width=0.24\textwidth]{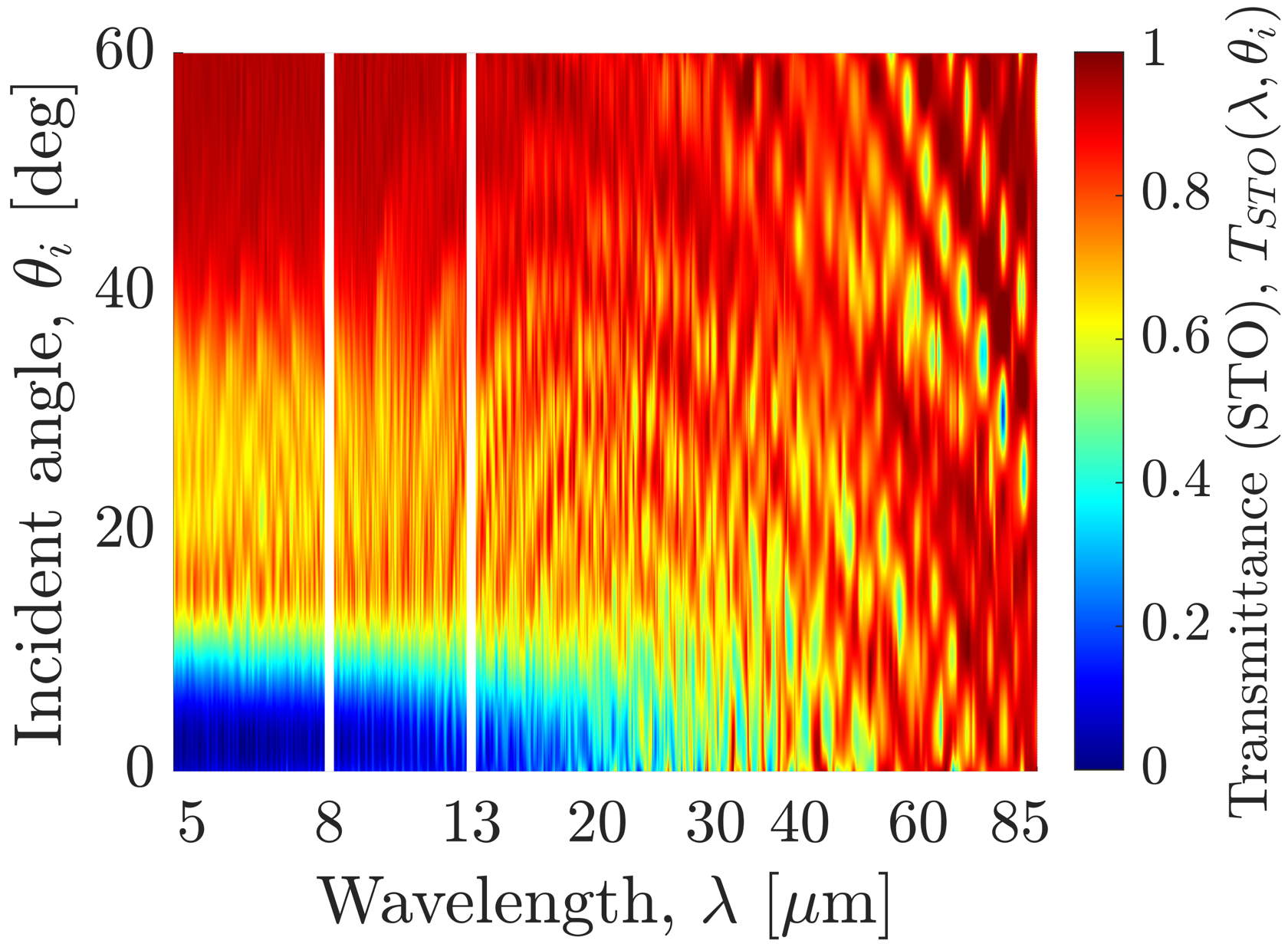}
    }
    \subfloat[Reflectance (OTS)\label{fig:2c}]{
        \includegraphics[width=0.24\textwidth]{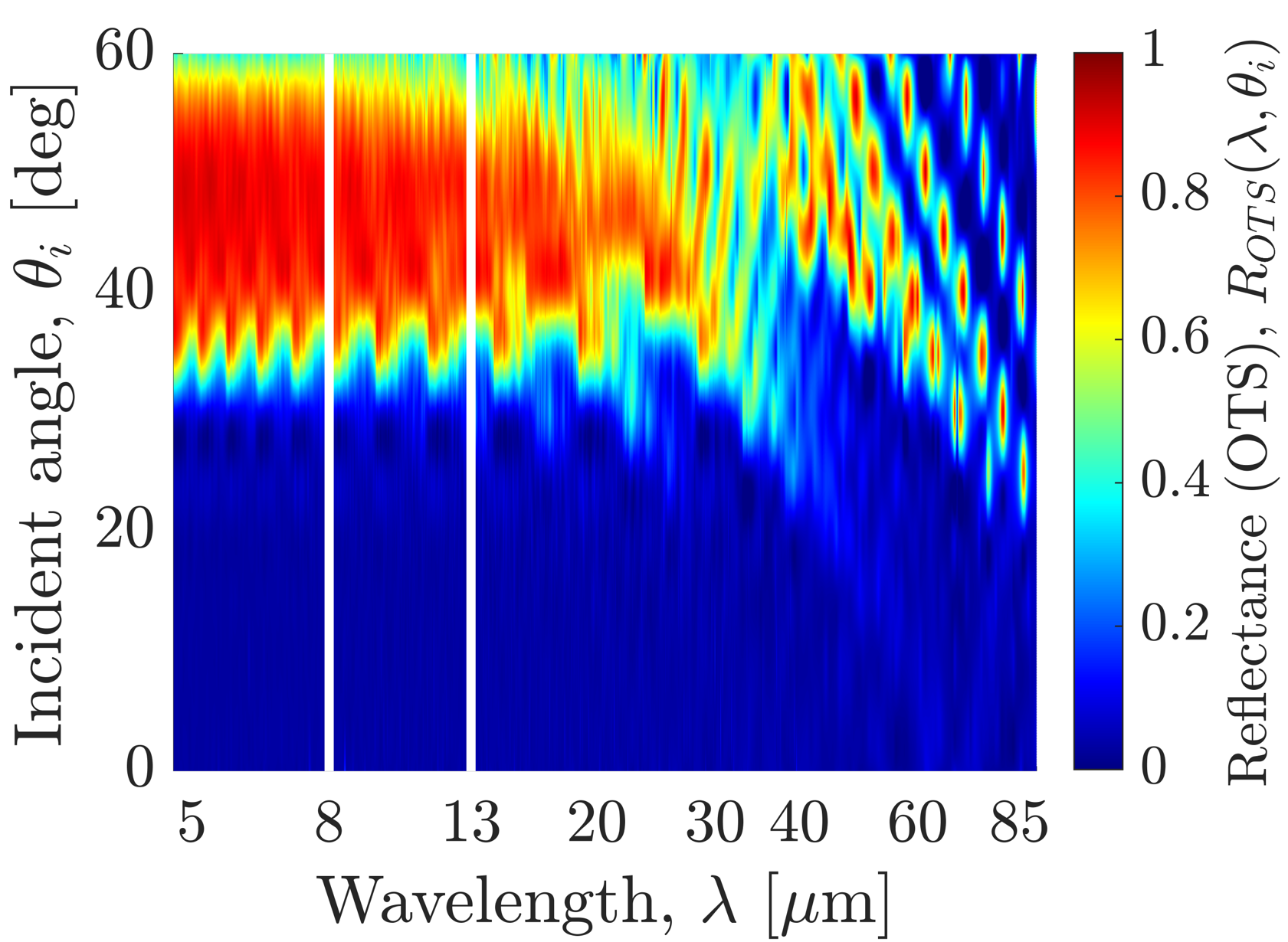}
    }
    \subfloat[Transmittance (OTS)\label{fig:2d}]{
        \includegraphics[width=0.24\textwidth]{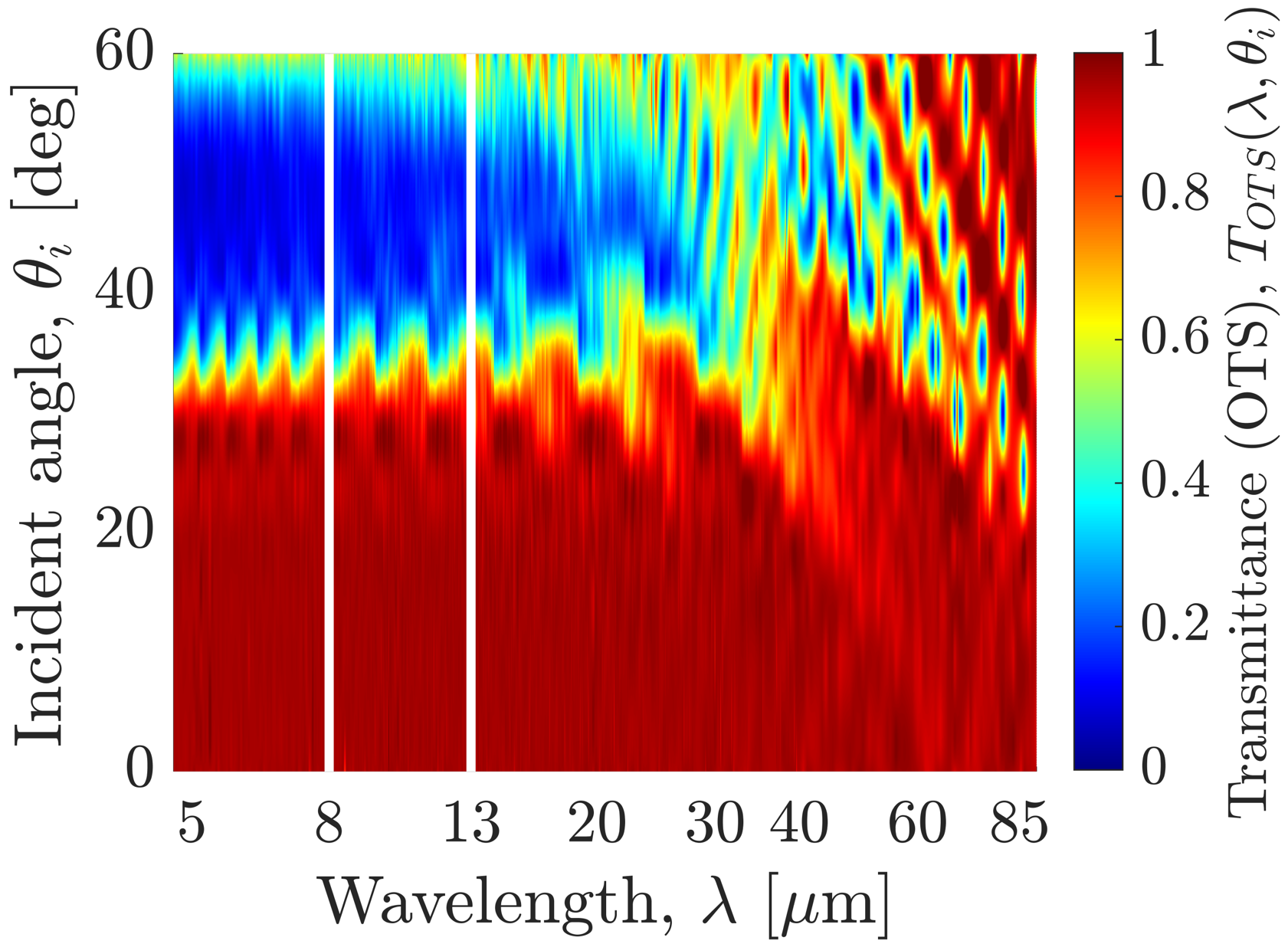}
    }

    \caption{Angle-wavelength maps of the reflection and transmission coefficients for sky-to-object and object-to-sky incidence: (a) $R_{\mathrm{STO}}(\lambda,\theta_i)$, (b) $T_{\mathrm{STO}}(\lambda,\theta_i)$, (c) $R_{\mathrm{OTS}}(\lambda,\theta_i)$, and (d) $T_{\mathrm{OTS}}(\lambda,\theta_i)$. The pronounced normal-incidence asymmetry progressively degrades at oblique angles, with increased STO transmission and reduced OTS transmission over a broad spectral range.}

    \label{fig:oblique}
\end{figure*}

\section{Full-Wave Electromagnetic Analysis}
\label{sec:overview_DEC}

This section presents a two-dimensional full-wave EM solver based on the DEC for evaluating the reflection and transmission of obliquely incident plane waves from a periodic PRC unit cell.
The structure is assumed to be periodic along the $x$ direction and illuminated by a TE$^{z}$-polarized plane wave.

\subsection{Problem description and governing equation}
\label{subsec:governing_eqn}

We consider a two-dimensional unit cell of size $L_x \times L_y$ containing an inhomogeneous, non-magnetic dielectric medium with permittivity $\epsilon(\mathbf{r})=\epsilon_r(\mathbf{r})\epsilon_0$, embedded in free space.
The structure is periodic along the $x$ direction, and the computational domain is denoted by $\Omega$.
Unless otherwise specified, a dispersionless dielectric constant $\epsilon_r=2.56$ is assumed.
A TE$^{z}$-polarized plane wave is incident on the unit cell at an angle $\theta_i$, with incident electric field
$\mathbf{E}^{(\mathrm{i})}(\mathbf{r})=\mathbf{E}_0 \exp\!\left(i \mathbf{k} \cdot \mathbf{r} \right)$,
where the wave vector $\mathbf{k}=(k_x,k_y)$ satisfies $|\mathbf{k}|=k_0=\omega\sqrt{\mu_0\varepsilon_0}$ and $\theta_i=\tan^{-1}(|k_y|/|k_x|)$.
Adopting a scattered-field formulation, the scattered electric field $\mathbf{E}^{(\mathrm{s})}$ satisfies
\begin{equation}
\left[
\nabla \times \mu_0^{-1} \nabla \times
-
\omega^2 \epsilon(\mathbf{r})
\right]
\mathbf{E}^{(\mathrm{s})}(\mathbf{r})
=
\omega^2 \chi_e(\mathbf{r})\,\mathbf{E}^{(\mathrm{i})}(\mathbf{r}),
\label{eq:scattered_pde}
\end{equation}
where $\chi_e(\mathbf{r})$ denotes the electric susceptibility.
The total electric field is given by
$\mathbf{E}^{(\mathrm{t})}=\mathbf{E}^{(\mathrm{i})}+\mathbf{E}^{(\mathrm{s})}$.

\subsection{DEC discretization}

We discretize \eqref{eq:scattered_pde} using the DEC framework.
The scattered electric field is expanded in terms of Whitney 1-forms~\cite{Bossavit1988} as
$\mathbf{E}^{(\mathrm{s})}(\mathbf{r})=\sum_{i=1}^{N_e} e_i\,\mathbf{W}_i^{(1)}(\mathbf{r})$ where $e_i$ denotes the line-integral degree of freedom associated with the $i$th mesh edge.
Applying DEC yields the sparse linear system
\begin{equation}
\left(
\overline{\mathbf{C}}^{T}\cdot [\star_{\mu^{-1}}]\cdot \overline{\mathbf{C}}
-
\omega^2[\star_{\epsilon}]
\right)\cdot\mathbf{E}
=
\mathbf f,
\label{eqn:DEC}
\end{equation}
where $\overline{\mathbf{C}}$ is the incidence matrix representing the discrete curl operator.
The discrete Hodge matrices and source vector are defined as
$[\star_{\epsilon}]_{ij}
=
\int_{\Omega}
\epsilon(\mathbf{r})\,
\mathbf{W}_i^{(1)}\cdot
\mathbf{W}_j^{(1)}\,dV
$, $[\star_{\mu^{-1}}]_{ij}
=
\int_{\Omega}
\mu_0^{-1}\,
\mathbf{W}_i^{(2)}\cdot
\mathbf{W}_j^{(2)}\,dV
$, and $[\mathbf f]_i
=
\int_{\Omega}
\omega^2\chi_e(\mathbf{r})\epsilon_0\,
\mathbf{E}^{(\mathrm i)}\cdot
\mathbf{W}_i^{(1)}\,dV
$.

\subsection{Bloch periodic boundary conditions}

Lateral periodicity is enforced by Bloch PBC on the $\pm x$ boundaries.
For an obliquely incident plane wave with in-plane wavevector component $k_x=k_0\sin\theta_i$, the electric field satisfies $\mathbf{E}(x+L_x,y)=e^{ik_xL_x}\mathbf{E}(x,y)$.
In the DEC formulation, corresponding edge-based degrees of freedom on the left and right boundaries are related by a complex phase factor, eliminating redundant unknowns and yielding a reduced system that implicitly incorporates the Bloch phase delay.
This enables efficient angle-resolved simulations using a single periodic unit cell.

\subsection{Floquet reflection and transmission coefficients}
\label{sec:floquet_RT}

Under Bloch-periodic excitation, the EM fields in the homogeneous regions are expanded in Floquet harmonics indexed by the diffraction order $m$.
For each order, the wavevector components are $k_x^{(m)} = k_x + \frac{2\pi m}{L_x}$ and $k_y^{(m)} = \sqrt{k_0^2-\left(k_x^{(m)}\right)^2}$ where only propagating modes with real-valued $k_y^{(m)}$ are considered.
The reflection and transmission coefficients are obtained by projecting the computed fields onto the corresponding Floquet modes at reference planes,
\begin{align}
R_m &=
\frac{1}{L_x}
\int_0^{L_x}
E_x^{(\mathrm{s})}(x,y_{\mathrm{top}})
e^{-ik_x^{(m)}x}\,dx,
\\
T_m &=
\frac{1}{L_x}
\int_0^{L_x}
E_x^{(\mathrm{t})}(x,y_{\mathrm{bot}})
e^{-ik_x^{(m)}x}\,dx.
\end{align}
The total reflected and transmitted power coefficients are evaluated as $R=\sum_{m\in\mathcal P}\frac{k_y^{(m)}}{k_y^{(0)}}|R_m|^2$ and $T=\sum_{m\in\mathcal P}\frac{k_y^{(m)}}{k_y^{(0)}}|T_m|^2$, respectively, where the factor $k_y^{(m)}/k_y^{(0)}$ accounts for the normal Poynting-flux ratio.

\begin{figure}[t]
    \centering
    \subfloat[Normal incidence\label{fig:2a}]{
        \includegraphics[width=0.4125\textwidth]{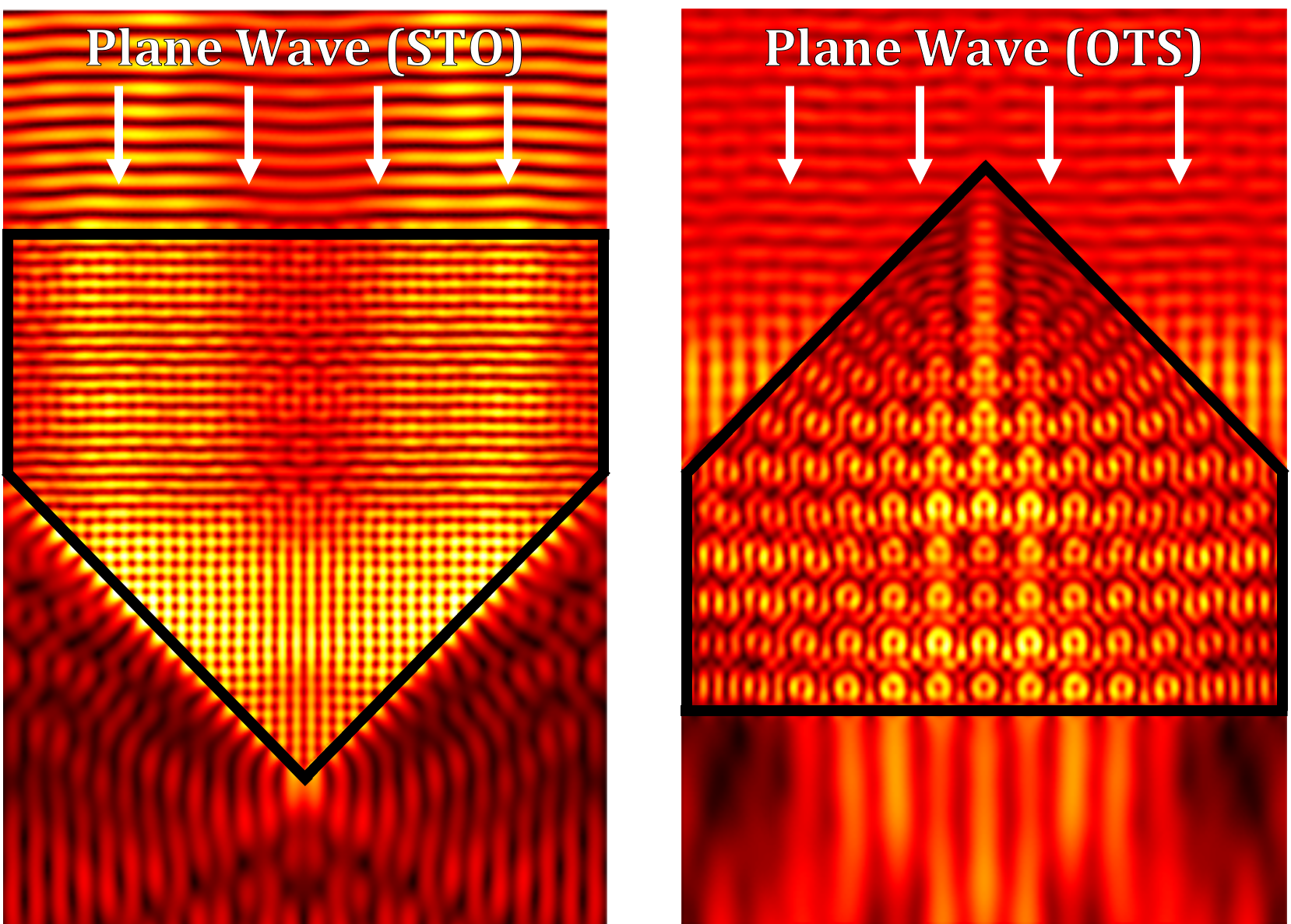}
    }        
    \\
    \subfloat[Oblique incidence\label{fig:2b}]{
        \includegraphics[width=0.4125\textwidth]{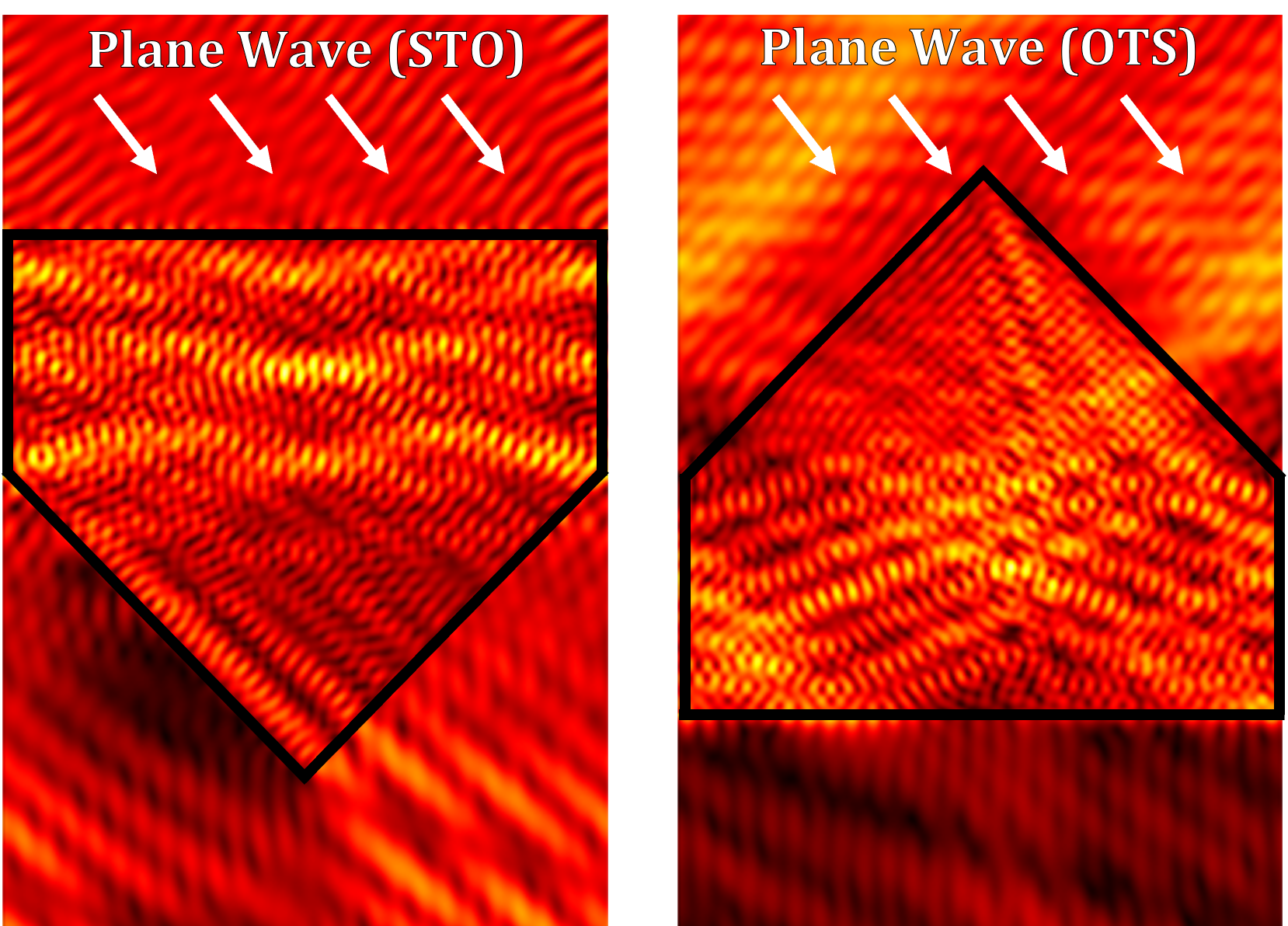}
    }    
    \caption{Spatial distributions of the electric-field magnitude $|\mathbf{E}|$ at a wavelength of $10~\mu$m: (a) normal incidence for STO and OTS, and (b) oblique incidence ($50^\circ$) for STO and OTS. The results highlight the distinct transmission behaviors of STO- and OTS-based structures at grazing incidence.}
\end{figure}

\section{Energy-Balance-Based Thermal Analysis}

We consider a non-contact transmissive PRC configuration in which a thermally emitting object is placed behind an engineered IR-transmissive layer with direction-dependent optical response.
Solar irradiation is neglected, corresponding to nighttime operation or ideal solar rejection, and the surrounding environment is modeled as an isotropic thermal radiation field at ambient temperature $T_{\mathrm{amb}}$.

The blackbody spectral radiance at temperature $T$ is given by Planck’s law,
$
I_{\mathrm{bb}}(\lambda,T)
=
\frac{2hc^2}{\lambda^5}
\left[\exp\!\left(\frac{hc}{\lambda k_{\mathrm{B}}T}\right)-1\right]^{-1}.
$
For a planar interface, radiative power exchange is evaluated through hemispherical angular integration weighted by the projected-area factor $\cos\theta$.
Accordingly, the radiative power exchanged through a half space is written as
\begin{equation}
\frac{\mathcal{P}(T;q)}{2\pi}
=
\!\int_0^{\theta_{\max}}\!\cos\theta\sin\theta
\left[\int_{\lambda_{\text{min}}}^{\lambda_{\text{max}}}
q(\lambda,\theta)\,I_{\mathrm{bb}}(\lambda,T)\,d\lambda\right]d\theta,
\end{equation}
where $q(\lambda,\theta)$ is a dimensionless spectral--angular weighting coefficient ($0\le q\le 1$) that accounts for the direction-dependent optical response of the intermediate PRC layer, and $\theta_{\max}=60^\circ$ is chosen to match the available angular scattering data.
The wavelength limits $\lambda_{\text{min}}$ and $\lambda_{\text{max}}$ define the finite spectral bandwidth over which the EM response of the PRC structure is evaluated, corresponding to the wavelength range resolved in the full-wave simulations and encompassing the dominant thermal emission band.
The PRC layer is characterized by direction-dependent spectral transmittance, denoted by $\tau_{\mathrm{OTS}}(\lambda,\theta)$ for the object-to-sky direction and $\tau_{\mathrm{STO}}(\lambda,\theta)$ for the sky-to-object direction.
The radiative power incident on the object from the ambient and the radiative power emitted by the object to the environment are given by $P_{\mathrm{in}}(T_{\mathrm{amb}})
=
\mathcal{P}\!\left(T_{\mathrm{amb}};\tau_{\mathrm{STO}}\right)$ and $P_{\mathrm{out}}(T)
=
\mathcal{P}\!\left(T;\tau_{\mathrm{OTS}}\right)
$.
To elucidate the impact of angular selectivity, two modeling assumptions are compared.
In Case~A, the normal-incidence transmittance is applied uniformly to all angles, i.e., $\tau(\lambda,\theta)\equiv\tau(\lambda,0)$.
In Case~B, the full angular-dependent transmittance $\tau(\lambda,\theta)$ obtained from EM simulations is directly incorporated into the hemispherical integration.
The object and any thermally well-coupled substrate are modeled as a lumped thermal system with effective areal heat capacity $C$.
The net heat flux into the object is
\begin{equation}
P_{\mathrm{net}}(T)
=
P_{\mathrm{in}}-P_{\mathrm{out}}(T)-h_c\!\left(T-T_{\mathrm{amb}}\right),
\end{equation}
where $h_c$ denotes the effective convective (and conductive) heat-transfer coefficient.
The transient temperature evolution follows
\begin{equation}
C\frac{dT}{dt}=P_{\mathrm{net}}(T),
\label{eq:energy_balance_ode}
\end{equation}
which is integrated numerically from the initial condition $T(0)=T_0$.
The long-time limit of the time-domain integration defines the dynamically stable steady-state temperature $T_\infty$.

\section{Numerical Results}
\label{sec:RESULTS}

\subsection{DEC EM Solver validation}
To verify the accuracy of the in-house DEC solver, reflectance and transmittance spectra computed for the same unit cell under normal incidence are compared with FDTD results, as shown in Fig.~\ref{fig:femfdtd}.
Here, $R_{\mathrm{STO}}(\lambda,\theta_i)$ denotes the power reflectance for STO incidence at free-space wavelength $\lambda$ and incidence angle $\theta_i$. The corresponding OTS quantity is denoted by $R_{\mathrm{OTS}}(\lambda,\theta_i)$.
Both methods reproduce the overall spectral trends and resonance features of $R_{\mathrm{STO}}(\lambda,0^\circ)$ and $T_{\mathrm{STO}}(\lambda,0^\circ)$ across the entire wavelength range.
In particular, high reflection and low transmission are observed at short wavelengths, followed by a gradual decrease in reflection and an increase in transmission as the wavelength increases.
This confirms that both solvers correctly capture the intended suppression of sky-to-object transport under normal incidence.
Since the structure is designed to be IR-transparent and effectively lossless, the absorbed power is negligible in the wavelength band of interest, so the reflected and transmitted power fractions are expected to be complementary up to numerical error.
Both solvers exhibit complementary variations of reflection and transmission over most of the spectrum, satisfying this condition.
Minor discrepancies in limited wavelength ranges are attributed to differences in mesh resolution, PML settings, Floquet mode truncation, and numerical integration.
Based on this comparison, the in-house DEC solver is validated and subsequently applied to oblique-incidence analysis and angle-resolved thermal modeling.

\begin{figure}[t]
\centering
\includegraphics[width=.9\linewidth]{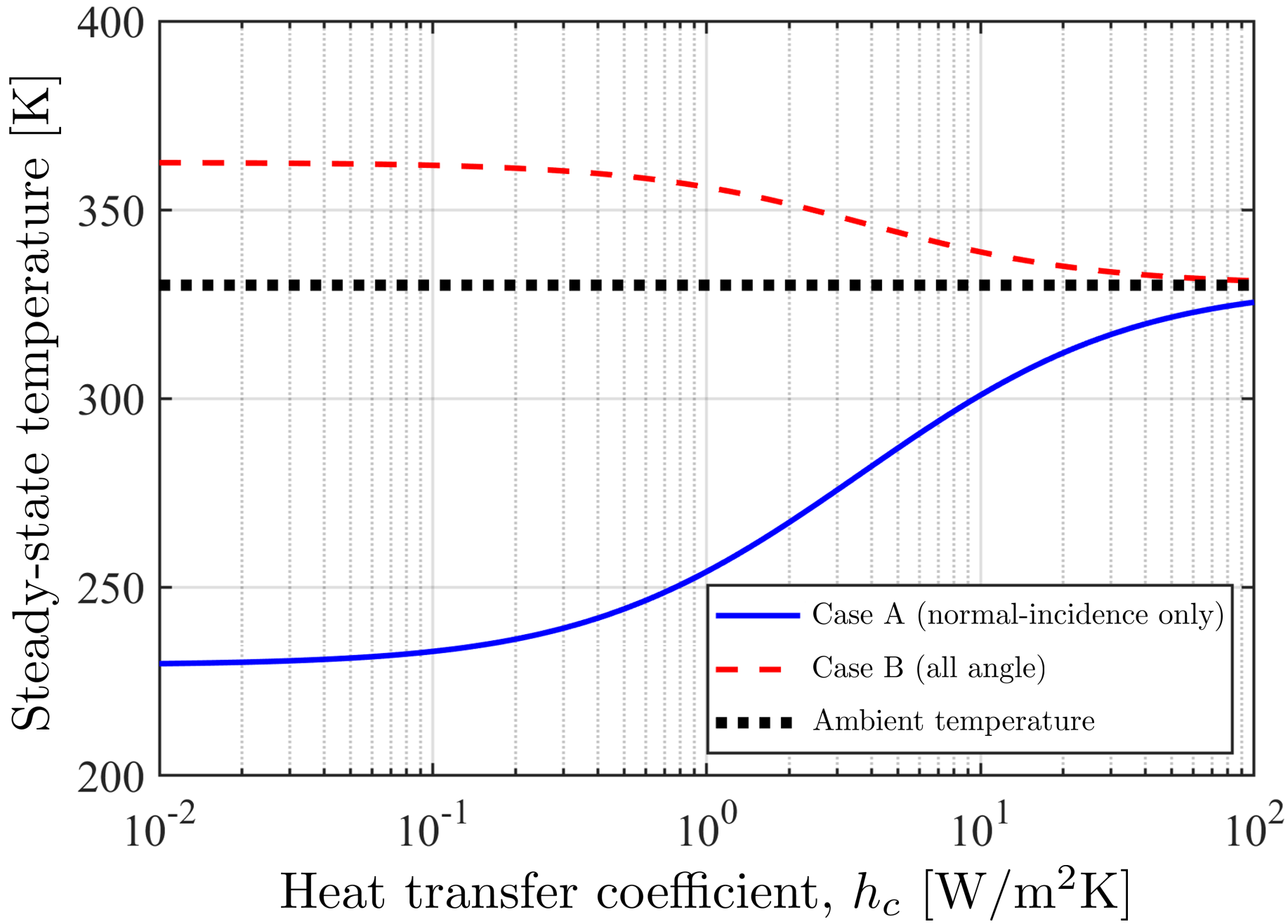}
\caption{
Steady-state temperature distributions as a function of the effective heat transfer coefficient, comparing cases where only normal-incidence transmittance is considered and where transmittance over all incidence angles is taken into account.
Here, $T_\infty$ is the steady-state object temperature predicted by \eqref{eq:energy_balance_ode}, $T_{\mathrm{amb}}$ is the ambient temperature, and $h_c$ (W\,m$^{-2}$\,K$^{-1}$) is the effective convective (and conductive) heat-transfer coefficient between the object and ambient air.}
\label{fig:temperature}
\end{figure}

\subsection{Angle--wavelength dependence of $R$ and $T$}
Fig.~\ref{fig:oblique} shows the angle--wavelength distributions of reflectance and transmittance for both STO and OTS directions.
Near normal incidence, pronounced asymmetric transport is observed in the MIR region: $T_{\mathrm{STO}}(\lambda,\theta_i)$ remains low while $R_{\mathrm{STO}}(\lambda,\theta_i)$ is high, whereas $T_{\mathrm{OTS}}(\lambda,\theta_i)$ is significantly enhanced, enabling efficient outward radiation from the object.
As the incidence angle increases, this asymmetry rapidly degrades.
Beyond a modest deviation from normal incidence (approximately $\theta_i \gtrsim 10^\circ$), STO transmission increases while OTS transmission decreases, effectively reversing the intended transport direction.
Importantly, this reversal persists over a broad angular range rather than being confined to a narrow region.
These results indicate that reliance on normal-incidence spectra alone can substantially overestimate net radiative cooling and highlight the necessity of angularly resolved integration of $R$ and $T$.

\subsection{Steady-state temperature prediction}
Fig.~\ref{fig:temperature} shows the predicted steady-state temperature $T_\infty$ as a function of the effective convective (and conductive) heat-transfer coefficient $h_c$, obtained from the energy-balance model in \eqref{eq:energy_balance_ode}.
The steady-state temperature is defined as the long-time limit of the time-domain integration starting from $T(0)=T_0$.
To assess the impact of angular selectivity, two cases are compared.
Case~A applies the normal-incidence transmittance uniformly to all angles, whereas Case~B incorporates the full angle--wavelength dependent transmittance obtained from EM simulations.
In Case~A, the model predicts $T_\infty < T_{\mathrm{amb}}$ over a wide range of $h_c$, suggesting effective cooling based on normal-incidence asymmetry.
In contrast, Case~B yields $T_\infty \gtrsim T_{\mathrm{amb}}$, indicating that cooling is suppressed once the full angular distribution is taken into account.
This discrepancy is consistent with Fig.~\ref{fig:oblique}, where asymmetric MIR transport is confined to near-normal incidence and rapidly collapses at oblique angles.
As a result, hemispherical integration increases ambient-to-object radiative gain while reducing object-to-sky radiative loss, eliminating the net-cooling regime implied by normal-incidence-only models.
As $h_c$ increases, $T_\infty$ approaches $T_{\mathrm{amb}}$ in both cases due to enhanced non-radiative heat exchange.
Overall, these results confirm that practical IR-transparent PRC requires angularly distributed asymmetric transparency rather than single-angle asymmetry.

\newpage




\ifCLASSOPTIONcaptionsoff
\newpage
\fi

\bibliographystyle{IEEEtran}
\bibliography{IEEEabrv,bibfile}

\end{document}